2nd URSI AT-RASC, Gran Canaria, 28 May – 1 June 2018

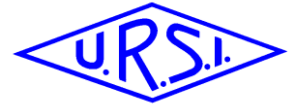

# Millimeter-Wave Over-the-Air Signal-to-Interference-plus-Noise-Ratio Measurements Using a MIMO Testbed

Koen Buisman*[1], David Cheadle[2], Tian Hong Loh[2], David Humphreys[2] and Thomas Eriksson[1]
(1) Dept. of Microtechnology and Nanoscience/ Dept. of Electrical Engineering, Chalmers University of Technology, Gothenburg, Sweden
(2) Engineering, Materials & Electrical Science Department, National Physical Laboratory (NPL), Teddington, United Kingdom

## Abstract

In this paper, over-the-air experiments with external and internal interferences were performed using Chalmers' millimeter-wave multiple-input-multiple-output testbed MATE. The resulting SINR for both interference experiments are compared and discussed.

## 1. Introduction

The high number of antenna elements and operation at millimeter-wave (mm-wave) frequencies will increase the system test-complexity [1-2] as there are many different kinds of hardware imperfections that could affect the quality of the received signals. In [2], a study on these imperfections has been carried out using different techniques. This paper presents some communications signals transmitted under interference conditions using the developed mm-wave multiple-input-multiple-output (MIMO) testbed, named MATE.

The paper is organized as follows: Section 2 introduces the testbed, section 3 contains the experimental results and discussion thereof. Section 4 concludes the paper.

## 2. MM-wave testbed MATE

The developed MATE testbed [1] operates between 28 – 31 GHz, with 1 GHz analog bandwidth per transmitter or receiver. MATE supports up to 18 channels, with up to 16 transmitters and up to 9 receivers. The experimental results in this paper are obtained using an 8 transmitter (TX), single receiver (RX) configuration (see Figure 1). Power limitations at the transmitter restrict our communication path to a ten centimeter over the air (OTA) channel. For future experiments over longer path lengths, amplifiers can be inserted.

The RF frontend is fully independent of the baseband hardware and software, thus can be replaced by other RF hardware, enabling different frequency bands. MATE is easily accessible to users by a remote access interface, for which a MATLAB client has been created, which takes care of all communication aspects, enabling worldwide access.

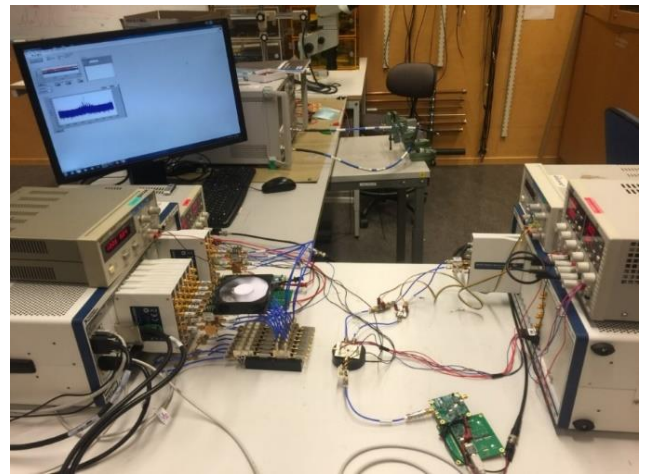

**Figure 1.** The MATE mm-wave testbed in 8x1 mode, showing the multi-channel transmitter on the left, and a single receiver on the right.

## 3. Experimental results

SINR experiments were performed, here we present measurement results for interference originating internal or external to the MATE testbed. All experiments were performed at a center frequency of 28.5 GHz.

### 3.1 External interference

When we have an external interferer in a multi-antenna system, the link quality will suffer. In the testbed, we simulate this case by using one of the antennas as an interferer, and the rest of the antennas operates as usual and transmits the communication signal in a beam towards the intended user. Then, by varying transmit powers, we can study the effect of more or less severe levels of interference. For the case of an external interferer, the interfering signal is assumed unsynchronized to the intended signal.

In Figure 2, we illustrate some resulting constellations, where we have varied the degree of interference. The distortion is quite Gaussian in its nature. In Figure 2(a), the measured results are dominated by the additive thermal noise from the receiver low noise amplifier (LNA), and for the other figures (i.e. Figure 2(b) and Figure 2(c)) the interference dominate.

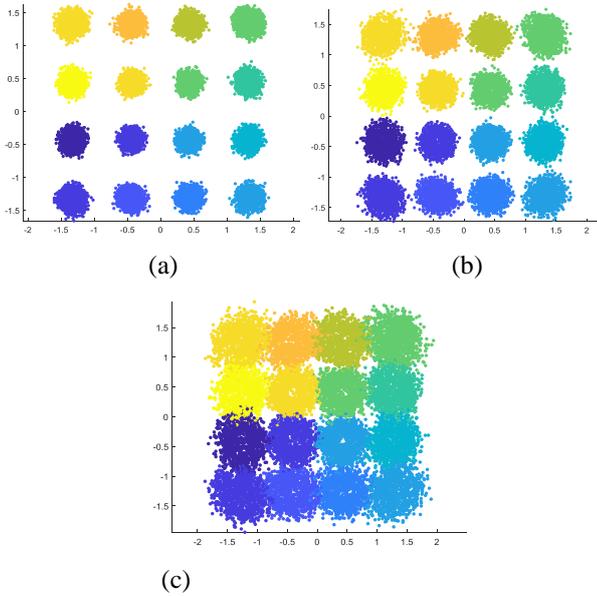

**Figure 2.** Constellations with three different SINR levels, 20, 16.5 and 13 dB respectively, for interference originating externally.

We also study the received SINR level when we vary the level of interference, in Figure 3. As can be expected, the SINR increases when we reduce the power of the interference, but we hit a ceiling when the thermal noise starts to dominate, so that the SINR never exceeds 29 dB in this experiment.

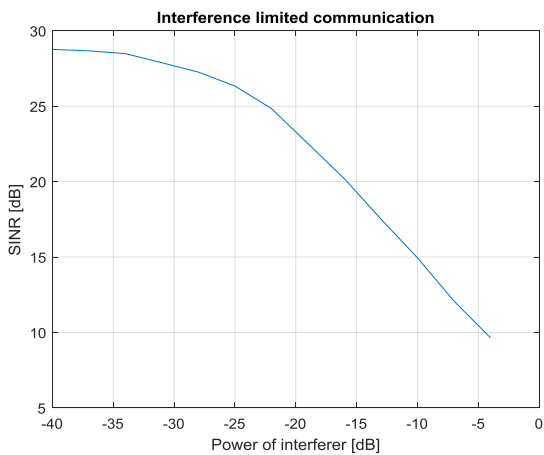

**Figure 3.** SINR as function of interferer relative power level.

## 3.2 Internal interference

A multiuser MIMO system, as will be the basis in 5G [], communicates to multiple users at the same time and at the same frequencies, differentiating between users only through spatial separation. If the channel-state information (CSI) is perfect, the users can be perfectly separated and no interference will occur. However, due to imperfect CSI acquisition, the users will interfere.

In this experiment, we have used the testbed to communicate to two users simultaneously, and varied the quality of the CSI to give various amounts of interference power. In the signal design, this corresponds to creating a precoding matrix with the intention to communicate with two users with known channel information. We consider one user as a reference user, and the other then becomes an interference source. The reference user transmits 16-QAM (quadrature amplitude modulation) symbols, while the interfering user transmits QPSK (quadrature phase shift keying) symbols. In Figure 4, we illustrate the received constellations at three different levels of interference.

We see that the distortion created is quite non-Gaussian; we can see some parts of the interfering user in the constellation of the studied user. This is typical for the in-system inter-user interference cases, since the users are synchronized within the system. As a consequence the noise is quite non-Gaussian.

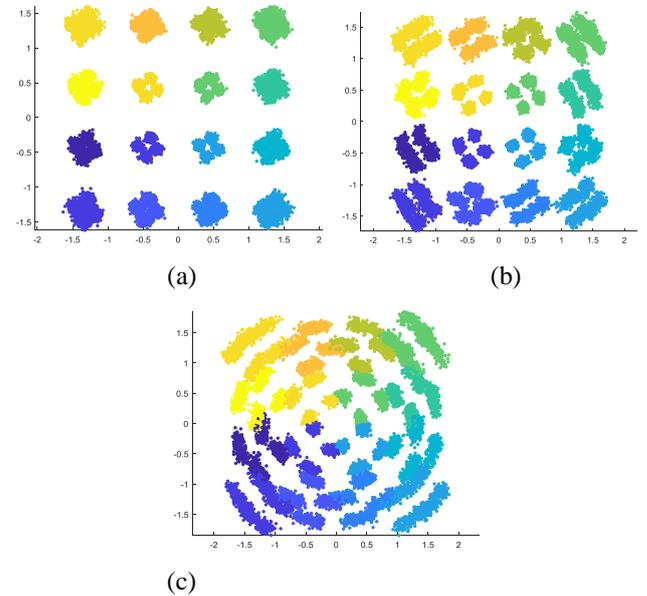

**Figure 4.** Constellations with three different SINR levels, 20, 16.5 and 13 dB respectively, for interference originating internally.

Typically, a given SINR leads to a certain symbol error probability. However, since the internal inter-user interference, with interference that is synced with the useful communication, has such a different probability density function when compared external interference, it could be expected that even for the same SINR the symbol errors could differ between the external and in-system cases. The peak-to-average of the interference is small in the synced case.
In Figure 5, we study the implications of this. Here, we have illustrated the symbol error rate for the in-system case and the external case, operating at the same levels of SINR.

The black curve with external interference behaves exactly as additive white Gaussian noise; the interference is quite Gaussian. However, the in-system interferer has a much less effect for small interference powers, and then a higher effect when the interference power is large (small SINR), leading to a steeper transition region. The effect is easily explained because of the non-Gaussian nature of the noise. For small values of the interference power, there will never be any symbol errors, since the received signal stays in the correct region always. Then, when the in-system interference is high, a lot of the symbols starts to be wrongly received.

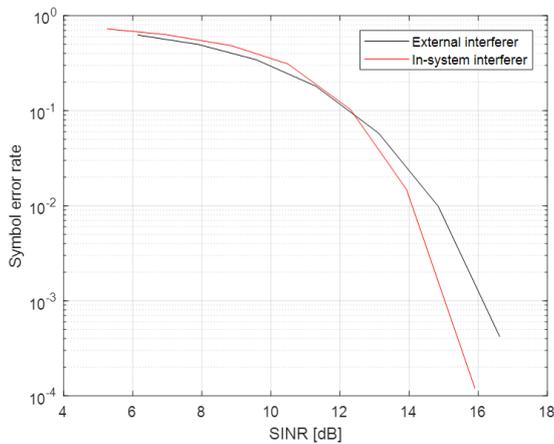

**Figure 5.** Comparison of Symbol error rate versus SINR between an external and internal interferer.

## 4. Conclusion

We have studied the case of external and internal inter-user interference in an mm-wave multi-antenna system, using the MATE testbed. The conclusion from these experiments are that the SINR is a quite good quality measure that predicts how good the performance of the system can be.

## 5. Acknowledgements

The results in this paper are the result of the project MET5G – Metrology for 5G communications. This project has received funding from the EMPIR programme co-financed by the Participating States and from the European Union's Horizon 2020 research and innovation programme.

## 6. References

1. K. Buisman and T. Eriksson, "Designing and Characterizing MATE, the Chalmers mm-wave MIMO Testbed," *accepted for publication in the 12th European Conference on Antennas and Propagation (EuCAP 2018) in Dec. 2017*.

2. S. Mumtaz, J. Rodriguez, and L. Dai, "mmWave Massive MIMO – A Paradigm for 5G," Elsevier Inc., 2017.